\documentclass[runningheads]{llncs}

\usepackage{amsmath}         
\usepackage{amssymb}         
\usepackage{graphicx}

\begin{document}

\title{Predictive Energy Management for Battery Electric Vehicles with Hybrid Models}

\author{Yu-Wen Huang\inst{1,2}\and
Christian Prehofer\inst{1}\and William Lindskog\inst{1}\and
 Ron Puts\inst{1}\and
 Pietro Mosca\inst{1}\and
Göran Kauermann\inst{2}
}
\authorrunning{Y. Huang et al.}

\institute{DENSO Automotive Germany \\
\email{c.prehofer@eu.denso.com} | \email{w.lindskog@eu.denso.com} | \email{r.puts@eu.denso.com} | \email{p.mosca@electravehicles.com\footnote{New affilitation}} \and
LMU München \\ \email{yu.huang@campus.lmu.de} | \email{goeran.kauermann@stat.uni-muenchen.de}}

\maketitle

\begin{abstract}
This paper addresses the problem of predicting the energy consumption for the
drivers of Battery electric vehicles (BEVs). Several external factors (e.g., weather) are shown to have huge impacts on the energy consumption of a vehicle besides the vehicle or powertrain dynamics. Thus,  it is challenging to take all of those influencing variables into consideration. 
The proposed approach is based on a hybrid model which improves the prediction accuracy of energy consumption of BEVs. The novelty of this approach is to combine a  physics-based simulation model, which captures the basic vehicle and powertrain dynamics, with a data-driven model. The latter accounts for other external influencing factors neglected by the physical simulation model, using 
machine learning techniques, such as generalized additive mixed models, random forests and boosting.
The hybrid modeling method is evaluated with a real data set from TUM and the hybrid models were shown that decrease the average prediction error from 40\% of the pure physics model to 10\%.


\end{abstract}

\section{Introduction}\label{intro}

Battery Electric Vehicles (BEVs) are fully electric vehicles with rechargeable batteries with the aim to reduce energy consumption and CO\textsubscript{2} emissions \cite{kongklaew2021barriers}. Compared to  vehicles with combustion engines, the heating needed to be provided by electrical energy, which can account for more than 30\% of the overall energy consumed \cite{steinstraeter2021effect}.
In addition, the battery capacity and efficiency are heavily affected by ambient temperature and external environmental conditions.
All of the reasons mentioned above lead to the problem of "range anxiety", which is the concern from the drivers that the car will not have enough energy to reach its final destination or the next recharging station. This is a significant barrier to market acceptance of battery electric vehicles (BEVs) \cite{pevec2019electric}.



The purpose of this paper is to utilize a hybrid modeling approach to improve the accuracy of energy consumption prediction. The novelty of this approach consists is the  combination of  a  physics-based model,  to capture the basic vehicle and powertrain dynamic, with a data-driven model (e.g., generalized additive mixed models), to account for the other potentially influencing factors that are not covered by the physics-based simulation model. 

Evaluation of the results is carried out using leave-one-out cross-validation. The data set used is from TU München \cite{steinstraeter2021effect}, which provides detailed trips' information of 72 trips driving around Munich in various ambient  driving conditions. 

    The contribution of the paper is twofold. First, the energy consumption prediction accuracy is improved despite the limited available trips in the data set and unknown number of drivers. Second, the proposed approach separates the vehicle specifics from the external influencing factors, which makes it easier to adapt the model to different vehicle, or to drivers with different driving behaviors.

\section{Background and Approach}\label{background}
\subsection{Prediction of Energy Consumption}\label{subsec:prediction_of_energy_consumption}
As the available battery capacity and energy consumption of the vehicle are the main influencing factors on the driving range of a BEV, a good deal of effort has been devoted to develop energy consumption estimation models for battery electric vehicles (e.g., \cite{halmeaho2017experimental}, \cite{ferreira2012data}). Many influencing factors are found that have impact on the energy consumption of BEVs, such as vehicle characteristics, vehicle speed, road elevation, acceleration, etc. These variables fluctuate greatly in real life, making energy consumption prediction an even more complex problem. 

In terms of physics-based simulation methodology, either Newton's Laws (analytical model) are applied to the vehicle as a point-mass, or detailed physical processes are incorporated for each module in an electric vehicle \cite{wu2019deep}. A number of analytical models were proposed in the literature, for instance, \cite{genikomsakis2017computationally} developed a simulation model, for energy estimation and route planning for a Nissan Leaf. \cite{halmeaho2017experimental} developed four simulation models for a city electric bus and
validated the models using the data collected from a bus prototype. The drawback of these purely physics-based models is that they don't consider heating and air conditioning in the energy computation, which can have huge impact on the overall consumption \cite{steinstraeter2021effect}.

In contrast to physics-based simulation models, a number of purely data-driven models (mostly regression models) were  proposed for the energy consumption estimation of BEVs. For instance, \cite{ferreira2012data} took into account many aspects (such as the battery's State of Charge (SOC), speed, weather data, road type, and driver profile) and gathered data from a BEV named Pure Mobility Buddy 09. The authors presented a data mining method for predicting a BEV's driving range that employs regression inside. \cite{de2017data} developed a cascade neural network (NN) regression model for energy consumption prediction. 

A recent paper \cite{petkevicius2021probabilistic} analyzed the same BEV tracking data in VED data set \cite{ved_dataset} to forecast SOC using deep-learning models, such as LSTM, and Deep Neural Networks. Yet this work does not consider the heating and AC energy.

Overall, data-driven models also come with shortcomings. For example, large amounts of data are required, and being more sensitive to the settings of hyper-parameters. Even though the prediction accuracy of the models can be high under certain circumstances, the result is not promising if a trip's characteristics are not seen in the training data set. In addition, the results are not physically justifiable due to the lack of interpretability of the black box models.

Therefore, a combination of the two above approaches is proposed. The goal of this work is to show that the proposed method, combining a physical model with a data-driven machine learning model, can be applied successfully and adapted to different driving behaviors and driving conditions.






\subsection{Hybrid Modeling for Energy Consumption Prediction}

\begin{figure}[!tb]
\centering
\includegraphics[width=\textwidth]{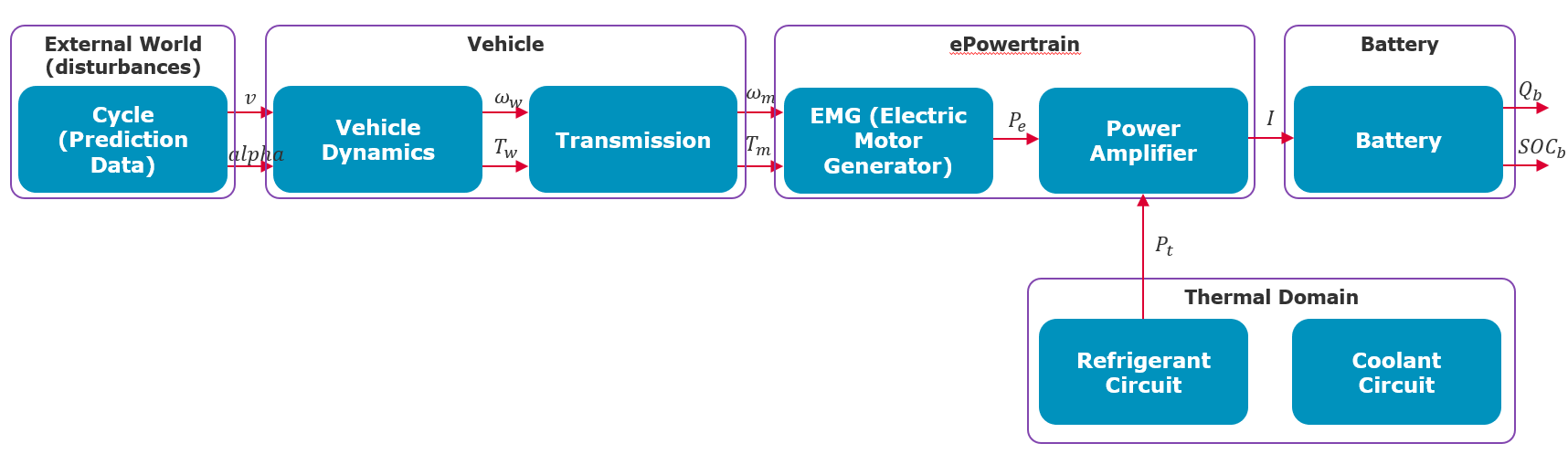}
\caption{Schematic view of utilized physic-based simulation model.}
 \label{fig:physical}
\end{figure}

Figure \ref{fig:physical} shows a schematic view of the physical model utilized physical model in this paper, following existing work as in \cite{guzzella2007vehicle}.
The physics-based simulation model is implemented in MATLAB  \cite{MATLAB:2017b} containing a Battery Electric Vehicle (BEV) model and its components such as motor, high voltage battery, and vehicle dynamics. Regenerative braking or  recuperation is the recovery of kinetic energy during braking. It is assumed to be 50\% as the data set contains more trips with higher ambient temperature without further tuning. 

\begin{figure}[!tb]
\centering
\includegraphics[width=10cm]{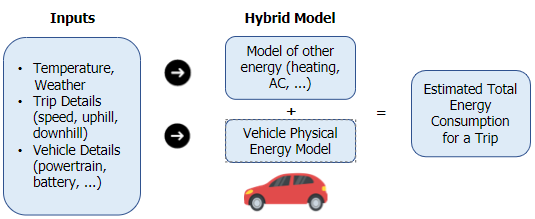}
\caption{Energy prediction for a trip utilizing proposed hybrid model.}
 \label{fig:hybrid}
\end{figure}

Figure  \ref{fig:hybrid} illustrates the proposed energy consumption prediction method for a specific trip. 
The physical model takes time, speed, and road inclination from the data as inputs, and calculates parameters containing information related to the battery of the vehicle, such as battery temperature, battery current, and battery voltage based on theories in physics. The whole physics simulation process is done in MATLAB.

For the statistical corrective models to account for the factors not considered by the physical model, longitudinal data models are used. Different methods have been proposed to handle longitudinal data effectively (thus eliminating the problem of dependence between temporal data), for example mixed-effects model \cite{fitzmaurice2012applied}. Each mixed model has two parts - fixed effects that are identical for the entire population and random effects which relate to each of the hierarchical levels. Fixed effects in our case describe general input variables e.g. temperature and time (duration of driving). 

Random effects, on the other hand, explain the randomness and heterogeneity as a result of known and unknown factors. In this work, this relates to trip specific heterogeneity which is not captured by the fixed effects. These are driver specific effects relating to both driving style and personal comfort temperature in the vehicle cabin. Mixed-effects models use local data to generate one broad flexible model for a certain area, explaining a significant portion of the population model's random variability \cite{adamec2019evaluation}.
In addition, this paper also utilized random forests and boosting, which are widely used in machine learning field for the comparison purpose.

The predictive capabilities of the models are evaluated using leave-one-out cross-validation. That is to say, there are a total of \(n\) iterations, where \(n\) is the number of the trips in the data set. For every iteration, there is only one testing trip, the remaining are all training data. The training data is used to train the statistical corrective models while the single testing trip is used to do an overall comprehensive evaluation. The evaluation process will repeat \(n\) times as there are \(n\) trips in the data set, and the calculation results will be averaged.

\subsection{Data Set}\label{datasets}

The TUM data set \cite{steinstraeter2021effect} contains 72 real driving excursions in Munich. The variables in the data are recorded once every 0.1 seconds using a 2014 BMW i3 (60 Ah) as the testing vehicle. Each trip has time-series data related to environmental signals (temperature, elevation, etc.), vehicle signals (speed, throttle), battery signals (voltage, current, temperature, SoC), heating circuit signals (indoor temperature, heating power) and a timestamp \cite{6jr9-5235-20}. Table \ref{tab:variables_tum} explains the relevant columns available in the TUM data set.

 \begin{table}[!tb]
\caption{Variable descriptions for TUM data set.}
\label{tab:variables_tum}
\centering
\begin{tabular}{ll}
\hline 
\textbf{Variable}       & \textbf{Description}\\
\hline
Time                 & Timestamp for each record in s\\
Trip.id                     & Trip identifier\\
Seasonality           & The trip is recorded in summer or winter\\
Weather                      & Weather when the trip was recording, e.g., cloudy\\
Velocity                     & The magnitude of instantaneous velocity in km/h\\
Elevation & The vehicle's height above sea level in m\\
Battery temperature                  & Battery temperature in $^{\circ}$C\\
Requested heating power                     & Heating power requested by the driver in kW\\
Air conditioner power                     & Air conditioner power used in kW\\
Ambient temperature                    & Outside temperature in $^{\circ}$C\\
Battery current                     & Battery current in A\\
Battery voltage                    &  Battery voltage in V\\
\hline
\end{tabular}
\end{table}


\section{Evaluation and Results}


This section assesses the results of proposed method quantitatively and the overall performance is discussed. The evaluation has been carried out on an ASUS VivoBook, Intel Core i7-8565U, 4 cores, 16GB RAM. The evaluation process can be summarized below:

\begin{enumerate}
    \item Calculate physical model energy consumption based on battery's current and voltage simulation results, and add the values as additional column to the trips in the data set.
    \item Subtract the real energy consumption by the physical model's predictions (from Step 1), which are named physical model's residuals.
    \item Build data-driven models to predict the physical model's residuals calculated.
    \item Add the estimated physical model's residuals from the data-driven models to the physical model's predictions from Step 1.
\end{enumerate}

To account for different scales, absolute percentage error is utilized for the cumulative values of the energy consumption at time point \(T\) of a trip (i.e., the end time of a trip), and is defined as:
\begin{equation}
\textrm{APE}_T = \bigg|\frac{y_T - f_T}{y_T}\bigg|
\end{equation} where \(y_T\) is the real value and \(f_T\) is the forecast value at time \(T\).

\subsection{Generalized Additive Mixed Models (GAMMs)}
Generalized additive models are an extension of regression models that allow effects of input variables to be smooth. For generalized additive mixed models, which includes mixed effects allows one to account for unobserved heterogeneity \cite{Everitt2001}, e.g., due to individual driving behavior. Suppose that observations of the $(i)$th of $(n)$ units consist of an outcome variable $(y_i)$ and $(p)$ covariates $(x_i = (1, x_{i1},..., x_{ip})^T)$ associated with fixed effects and a $(q \times 1)$ vector of covariates $(z_i)$ associated with random effects. Given a $(q \times 1)$ vector $(b)$ of random effects, the observations $(y_i)$ are assumed to be conditionally independent with means $(\mathbb{E}(y_i|b) = \mu_i^b)$ and variances var$((y_i|b) = \phi m_i^{-1}\upsilon(\mu_i^b))$, where $(\upsilon(\cdot))$ is a specified variance function, $(m_i)$ is a prior weight (e.g., a binomial denominator) and $(\phi)$ is a scale parameter, and follow a generalized additive mixed model

\begin{equation}
g(\mu_i^b) = \beta_0 + f_1(x_{i1}) + ... + f_p(x_{ip}) + z_i^Tb
\end{equation}

where \(g(\cdot)\) is a monotonic differentiable link function, \(f_j(\cdot)\) is a centred twice-differentiable smooth function, the random effects \(b\) are assumed to be distributed as \(\mathcal{N}\{0, D(\theta)\}\) and \(\theta\) is a
\(c \times 1\) vector of variance components \cite{lin1999inference}.

The estimates that are based on data that show clear violations of key assumptions for GAMM should be treated with caution, though few papers (e.g., \cite{schielzeth2020robustness}) state that mixed model estimates were usually robust to violations of those assumptions (e.g., normality of random effects). 

As the data does contain outliers,  the distribution of the response (i.e., physical model's residuals) heavily-tailed. The Student’s t family for heavy-tailed data is applied. The density of the Student’s t family is given by: 

\begin{equation}
f(y_i) = \frac{\Gamma((\nu+1)/2)}{\Gamma(\nu/2)}\frac{1}{\sqrt{\nu\pi}\sigma} \left(1+\frac{1}{\nu}\left(\frac{y_i-g(\eta_i)}{\sigma}\right)^2\right)^{-(\nu+1)/2}
\end{equation}

\begin{figure}[!tb]
\centering
\includegraphics[width=11cm]{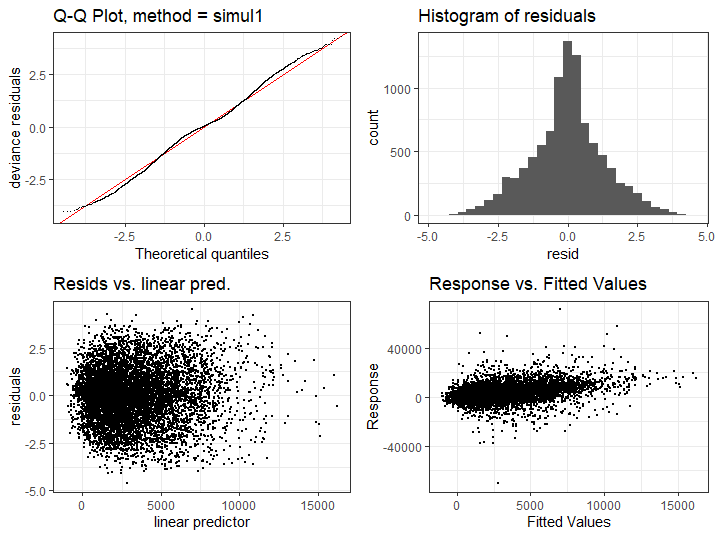}
\caption{Model diagnosis for GAMM with Student’s t family fitted on TUM data set.}
 \label{fig:check_tum_t}
\end{figure}

\begin{figure}[!tbp]
\centering
\includegraphics[width=\textwidth]{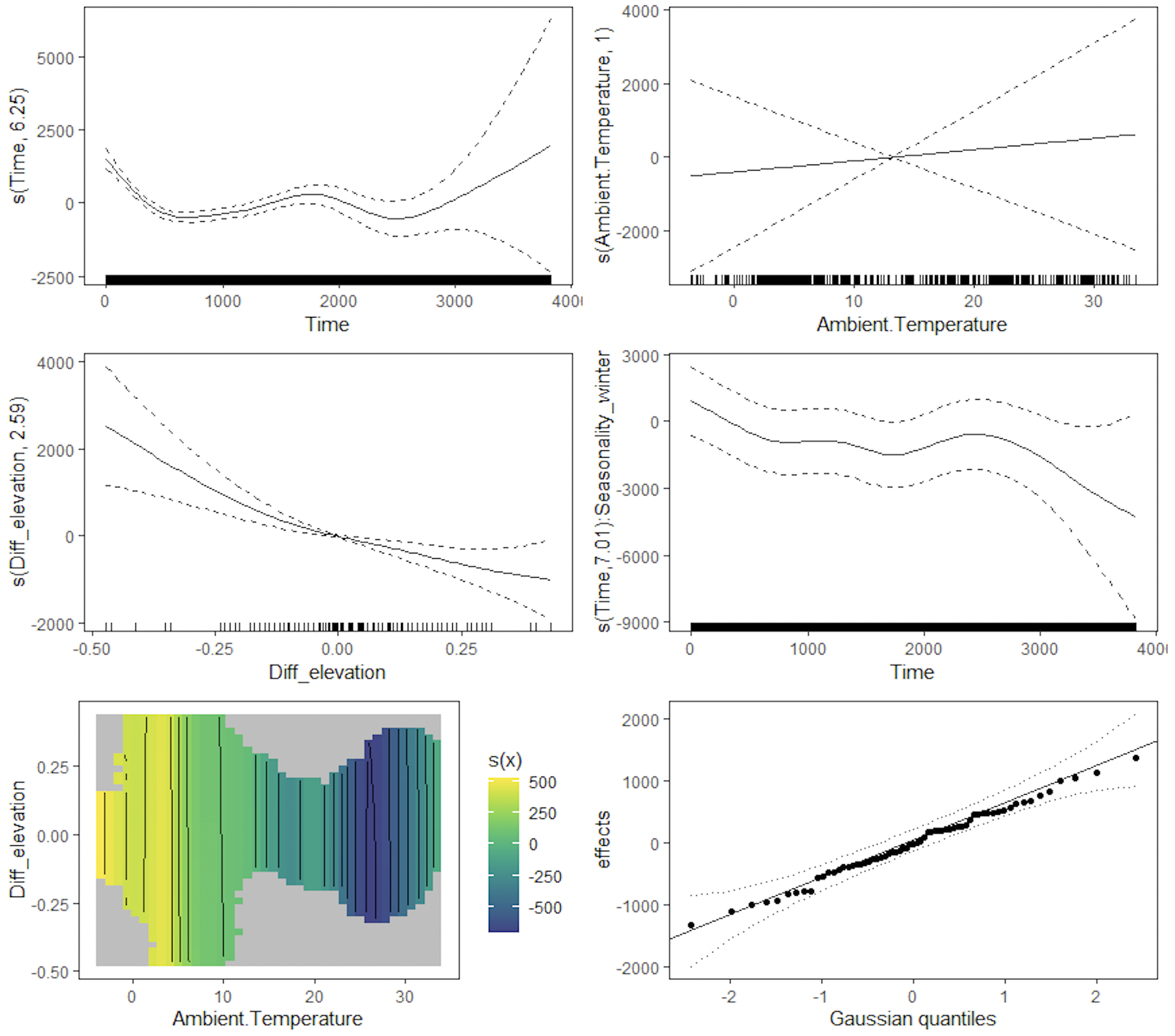}
\includegraphics[width=6cm]{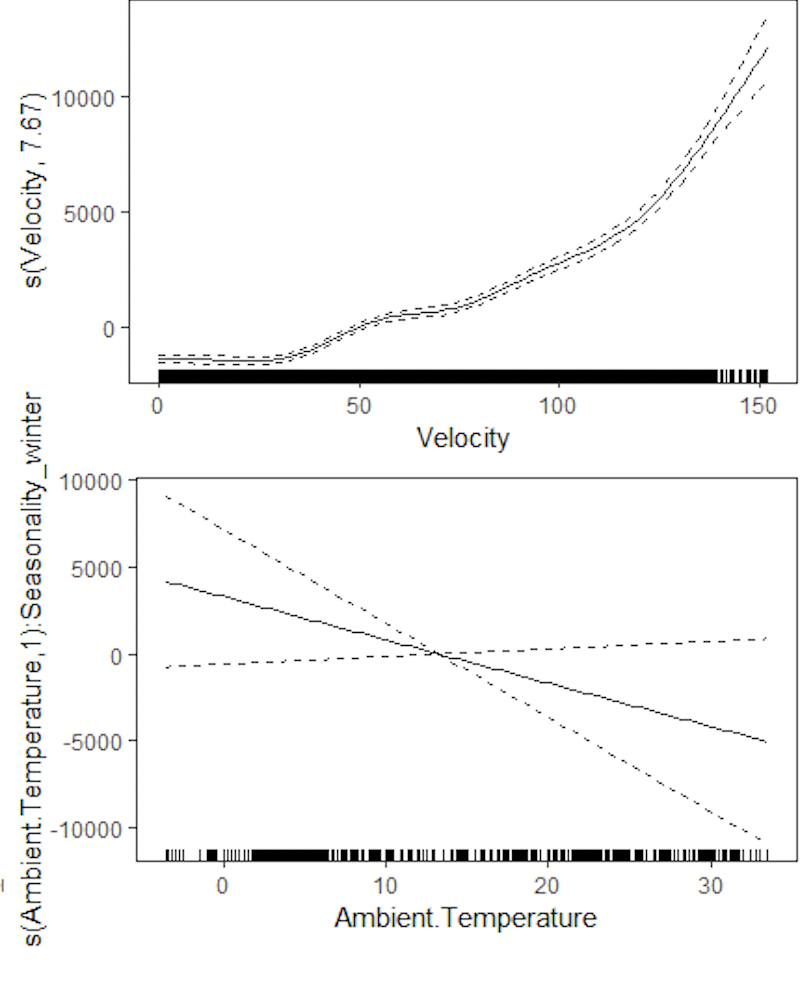}
\caption{Estimated smooths from the GAMM with Student’s t family fitted on the TUM data set.}
 \label{fig:smoothers_TUM_t}
\end{figure}

The model diagnosis for the GAMM with Student’s t family fitted on the TUM data set is shown in Figure \ref{fig:check_tum_t}. The Q-Q plot indicates that the Student’s t family is suitable for the TUM data set as the model's residuals from the GAMM match the theoretical quantiles (i.e., the red line). The estimated smooth effects from the GAMM model fit are presented in Figure \ref{fig:smoothers_TUM_t}. The last figure indicates the random effect "Trip Id" of the trips, and they are plotted against Gaussian quantiles. In mixed-effects models, one doesn't directly estimate random effects but assumes they are normally distributed with mean of 0. Therefore, the random effects are checked against Gaussian quantiles for the normally distributed assumption.

The summary of the GAMM model with Student’s t family fitted on the TUM data set is shown in Table \ref{table:coefficients_tum}. For the parametric terms of the model, although the p-values indicate that some of the predictors are not significant in terms of assuming threshold of 0.05, those predictors are still kept in the model as the main goal of the paper is to predict the energy consumption of a BEV instead of making explanation out of input variables (it makes little sense to use p-values to determine the variables in a model that is being used for prediction \cite{shmueli2010explain}). For the smooth terms of the model, such as time, the coefficient is not printed, as each smooth has several coefficients, one for each basis function. The Effective Degree of Freedom (EDF) represents the complexity of the smooth fit. EDF of 1 is equivalent to a straight line, EDF of 2 is equivalent to a quadratic curve, and so on. With higher EDF describing more wiggly curves. 

\begin{table}[!tb]
\begin{center}
\caption{Summary of GAMM with Student’s t family fitted on the TUM data set.}
\vspace*{5mm}
\begin{tabular}{l c}
\hline
 & Model TUM \\
\hline
(Intercept)                                     & $1537.44 \; (938.99)$     \\
Weather\_cloudy                                 & $796.09 \; (649.09)$      \\
Weather\_dark                                   & $1430.53 \; (723.35)^{*}$ \\
Weather\_dark\_little\_rainy                    & $0.00 \;   (0.00)$        \\
Weather\_rainy                                  & $0.00 \;   (0.00)$        \\
Weather\_slightly\_cloudy                       & $386.27 \; (582.56)$      \\
Weather\_sunny                                  & $457.96 \; (573.88)$      \\
Weather\_sunrise                                & $754.49 \; (751.99)$      \\
Weather\_sunset                                 & $2322.86 \; (998.28)^{*}$ \\
EDF: s(Time)                                    & $6.25 \;   (7.06)^{***}$  \\
EDF: s(Ambient.Temperature)                     & $1.00 \;   (1.00)$        \\
EDF: s(Velocity)                                & $7.67 \;   (8.54)^{***}$  \\
EDF: s(Diff\_elevation)                         & $2.59 \;   (3.35)^{***}$  \\
EDF: s(Time):Seasonality\_winter                & $7.01 \;   (7.92)^{***}$  \\
EDF: s(Ambient.Temperature):Seasonality\_winter & $1.00 \;   (1.00)$        \\
EDF: s(Ambient.Temperature,Diff\_elevation)     & $4.86 \;  (27.00)$        \\
EDF: s(Trip.id)                                 & $44.40 \;  (57.00)^{***}$ \\
\hline
\hline
\multicolumn{2}{l}{\scriptsize{$^{***}p<0.001$; $^{**}p<0.01$; $^{*}p<0.05$}}
\end{tabular}
\label{table:coefficients_tum}
\end{center}
\end{table}

\begin{figure}[!tb]
\centering
\includegraphics[width=10cm]{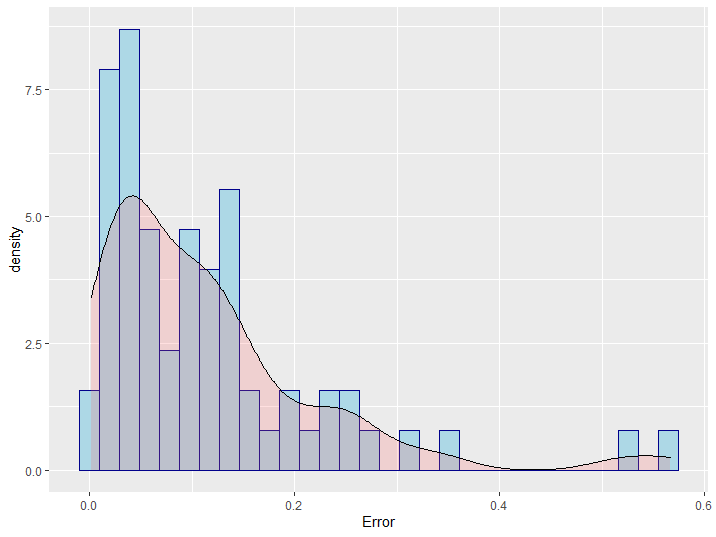}
\caption{Overall error distribution from LOOCV for GAMM with Student’s t family fitted on the TUM data set.}
 \label{fig:error_overall_tum_bam}
\end{figure}

The common way to evaluate a predictive model is by looking at the average error rate from all of the iterations of cross-validation, but it is still interesting to know how the model performs on each iteration. Figure \ref{fig:error_overall_tum_bam} shows the overall error distribution of all the iterations from the leave-one-trip-out cross-validation for the GAMM with Student’s t family fitted on the TUM data set.

\subsection{Random Forest}
For fitting Random Forest on longitudinal data, small modification on the normal random forest algorithm is proposed by \cite{karpievitch2009introspective} utilizing subject-level bootstrapping, which enables the effective use of all data samples and allows for unequal contribution from the subjects. That is to say, the algorithm grows each tree on a bootstrap sample (a random sample selected with replacement) at the subject level rather than at the replicate level of the training data. It has advantages for the analysis of longitudinal data because the sampling scheme is designed to accommodate data with multiple measurements for a given subject. Also, as the primary goal of the thesis is prediction, and not in performing inference on the model components, the method has shown to be effective in predicting cluster-correlated data. 

\begin{figure}[!tb]
\centering
\includegraphics[width=10cm]{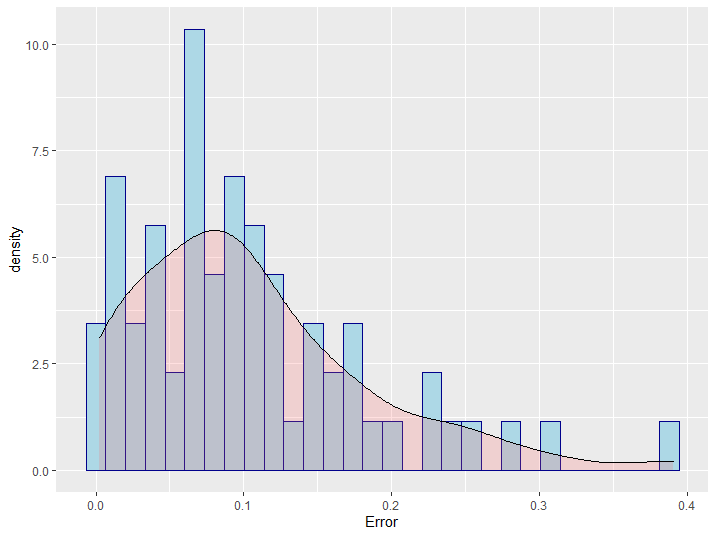}
\caption{Overall error distribution from LOOCV for random forest model fitted on the TUM data set.}
 \label{fig:error_overall_tum_rf}
\end{figure}

The random forest was fitted using the package htree in R, and the tuning of the hyper-parameters is done using grid search which attempts to compute the optimum values out of different combination of hyper-parameters. Figure \ref{fig:error_overall_tum_rf} shows the overall error distribution of all the iterations from the leave-one-trip-out cross-validation for the random forest model fitted on the TUM data set. 

\subsection{Boosting}
The boosting model utilized is using traditional gradient boosting method to form a marginal model as Generalized Estimating Equation (GEE). GEE, proposed by \cite{liang1986longitudinal}, is a general statistical approach to fit a marginal model for clustered data analysis. Whereas the mixed-effect model  is an individual-level approach by adopting random effects to capture the correlation between the observations of the same subject \cite{crowder1995use}, GEE is a population-level approach based on a quasi-likelihood function and provides the population-averaged estimates of the parameters \cite{wedderburn1974quasi}. In this method, the correlation between measurements is modeled by assuming a working correlation matrix. The estimations from the GEE are broadly valid estimates that approach the correct value with increasing sample size regardless of the choice of correlation model \cite{overall2004robustness}.

That is to say, if $(F_k)$ is the model at the $ (k)^{th}$ boosting iteration, a regression tree is fit to the residuals $(r_{ij} = y_{ij} - F_k(x_{ij}))$ for $(i = 1,...,n)$ and $(j = 1,...,n_i)$ (n subjects and $(i)$th subject observed at $(n_i)$ time points). If $(T_{k+1})$ denotes this regression tree, then the model is updated by $(F_{k+1}=F_{k}+T_{k+1})$, and the procedure is repeated. Determining the number of boosting iterations is done by using cross-validation with a leave-out-subject approach. 

\begin{figure}[!tb]
\centering
\includegraphics[width=10cm]{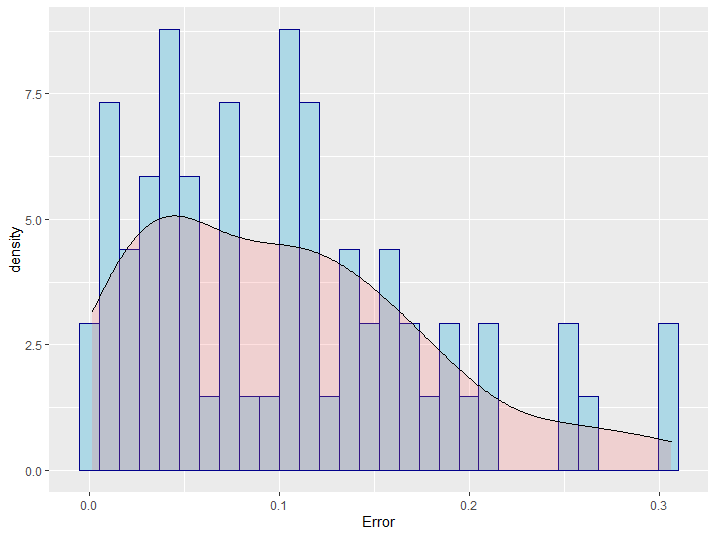}
\caption{Overall error distribution from LOOCV for boosting model fitted on the TUM data set.}
 \label{fig:error_overall_tum_tb}
\end{figure}

Figure \ref{fig:error_overall_tum_tb} shows the overall error distribution of all the iterations from the leave-one-out cross-validation for the boosting model fitted on the TUM data set. 

\subsection{Overall Comparison}
\begin{figure}[!tb]
\centering
\includegraphics[trim={0 0 0 0},clip,width=12cm]{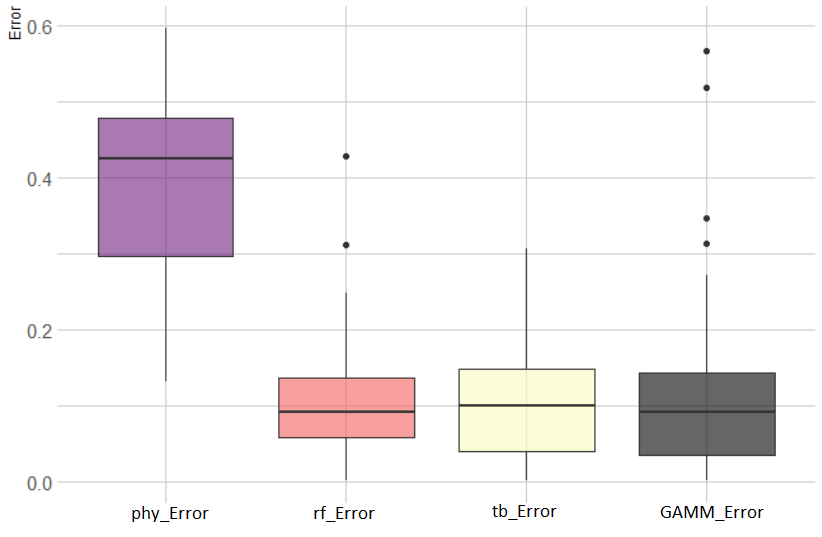}
\caption{Comparisons of the models fitted on the TUM data set.}
 \label{fig:overall_tum_model}
\end{figure}

The box plot for the errors from the models fitted on the TUM data set using Leave-one-out Cross-validation is shown in Figure \ref{fig:overall_tum_model}. Phy\_Error is the error from the physics-based simulation without utilizing hybrid modeling approach, Gamm\_Error is the error from the GAMM with Student’s t family hybrid modeling model, rf\_Error is the error from the random forest hybrid modeling model and tb\_Error is the error from the boosting hybrid modeling model. The average error rate for the models is shown in Table \ref{tab:avg_tum}.  

 \begin{table}[!h]
\caption{Average error rate from LOOCV for the models fitted on the TUM data set.}
\vspace*{5mm}
\label{tab:avg_tum}
\centering
\begin{tabular}{ll}
\hline 
\textbf{Model}       & \textbf{Average error} \\
\hline
Purely Physics-based model  &   0.379 \\
GAMM with Student’s t family  &  0.115  \\
Random forest   &   0.106\\
Boosting &  0.103\\

\hline
\end{tabular}
\end{table}

From the prediction accuracy perspective, three statistical corrective models don't show a large difference from each other. However, GAMMs have a longer running time compared to the others. In general, the shorter the running time, the better the model would be, as for the practical usage purpose, the drivers are typically not willing to wait for a long time to get a prediction of the cumulative energy consumption at the final destination. Therefore, ensemble learning models would be more preferable in this use case. GAMMs, on the other hand, are useful for observing the effects of each predictor.

\begin{figure}[!tb]
\centering
\includegraphics[trim={0 1cm 0 1cm},width=12cm]{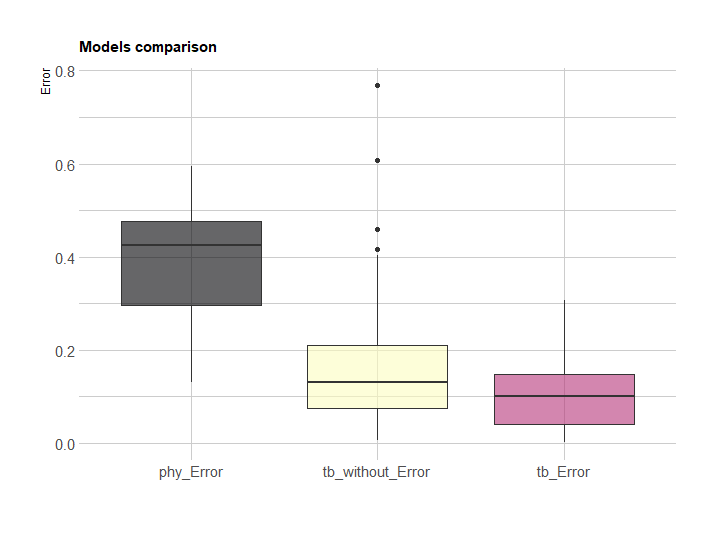}
\caption{Comparison of purely physics-based model, hybrid model and purely data-driven model fitted on the TUM data set.}
 \label{fig:benchmark_boost}
\end{figure}

In addition, as the boosting model has the best overall performance. The purely data-driven approach was constructed as a benchmark against hybrid modeling method utilizing boosting model. The same set of inputs are used to forecast real energy consumption instead of \(Residual_{Phy}\). The comparison of the error rate between the purely physics-based model's predictions, hybrid model using boosting model's predictions and purely data-driven boosting model's predictions is shown in Figure \ref{fig:benchmark_boost}. Phy\_Error is the error from the purely physics-based model without any statistical corrective models, tb\_without\_Error is the error from the purely data-driven approach using boosting model and tb\_Error is the error from the hybrid model using boosting, respectively. The purely physics-based model has the highest error, hybrid modeling approach has the lowest error among the three models, and the purely data-driven model has the highest variability compared to the other two, the maximum error rate is almost 0.8 (i.e., 80\%), which is really high in the real world. 

Although the data set comes with restrictions and limitations (e.g., unknown number of drivers, limited sample size issue, only one testing vehicle), the proposed hybrid modeling approach has shown to be very effective to a certain extent with the improvement of over 90\% of the total trips' energy consumption prediction accuracy utilizing either one of the corrective models. Despite the fact that the physics-based simulation model shows deviations to reality, it was corrected by the statistical models using less data than required by purely data-driven approaches. The idea of combining simplified physics-based models and data-driven models to achieve better prediction accuracy is therefore worth further investigation based on different scenarios.

\section{Conclusions}\label{conclusion}

Given the current deluge of sensor data and advances in statistical models or machine learning methods, the merging of principles from machine learning and physics will play an invaluable role in the future of scientific modeling to address the range anxiety of BEVs and physical modeling problems facing society. In this paper, a  physics-based model, representing a rather simple but very efficient model, has been successfully combined with corrective models into a hybrid prediction model to enhance its prediction capability under certain circumstances. While the physics-based model part captures what is actually known about the BEV, the statistical corrective models are responsible for describing the vehicle's excluded features and possible deviations from reality due to the simplified assumptions of the BEV model.  

On a realistic, challenging and small data set with a range of temperature and driving styles, the proposed method has shown to be able to achieve roughly 10\% average prediction error utilizing either one of the hybrid model.
Other works report same accuracy, e.g. \cite{petkevicius2021probabilistic}, however do not consider the effect of AC or heating at all, or another research on energy estimation \cite{holden2017development} also reports about 12\% error, however does not focus on e-vehicles where cabin (and battery) conditioning has a much stronger effect as it requires extra energy from the battery. 

Furthermore, the benefit of the proposed approach is the separation of the effect of powertrain dynamic and the other external factors, which makes it easier to apply trained model from the data obtained from one vehicle to another without sacrificing the prediction accuracy. Also, it can handle cases which are less frequently seen, but where the physical model has a strong effect, e.g. long downhill trips or cold/hot battery temperature.


\subsection*{Acknowledgments}
This work was partly supported by the TRANSACT project. TRANSACT (https://transact-ecsel.eu/) has received funding from the Electronic Component Systems for European Leadership Joint Under-taking under grant agreement no.  101007260.  This joint undertaking receives support from the European Union’s Horizon 2020 research and in-novation programme and Austria,  Belgium,  Denmark,  Finland,  Germany, Poland, Netherlands, Norway, and Spain.

\bibliography{main}
\bibliographystyle{splncs04}

\end{document}